\documentclass[superscriptaddress,twocolumn,aps,floatfix]{revtex4}
\usepackage{epsfig} 
\usepackage{amsmath} 

\begin{document}

\title{Network growth models and genetic regulatory networks}

\author{D.~V.~Foster} \affiliation{Physics Department and Center for
  Nonlinear and Complex Systems, Duke University, Durham, NC, 27514}
\author{S.~A.~Kauffman} \affiliation{Institute for Biocomplexity and Informatics, University of Calgary, Calgary, AB, Canada}
\author{J.~E.~S.~Socolar} \affiliation{Physics Department and Center
  for Nonlinear and Complex Systems, Duke University, Durham, NC,
  27514}

\date{\today}

\begin{abstract}
  We study a class of growth algorithms for directed graphs that are
  candidate models for the evolution of genetic regulatory networks.
  The algorithms involve partial duplication of nodes and their links,
  together with innovation of new links, allowing for the possibility
  that input and output links from a newly created node may have
  different probabilities of survival.  We find some counterintuitive
  trends as parameters are varied, including the broadening of
  indegree distribution when the probability for retaining input links
  is decreased.  We also find that both the scaling of transcription
  factors with genome size and the measured degree distributions for
  genes in yeast can be reproduced by the growth algorithm if and only
  if a special seed is used to initiate the process.
\end{abstract}


\maketitle

\section{Introduction}

The manufacturing by cells of the proteins necessary for sustaining
life is accomplished with the aid of molecular machinery that
translates DNA nucleotide sequences, or genes, into their specified
proteins at appropriate times.  The first step in making a protein is
the transcription process, in which the needed piece of RNA is formed
from the relevant gene.  The rate of transcription of a given gene can
often be enhanced or suppressed by the presence of proteins that bind
to the DNA near the gene, the gene's promoter region.  Once
transcription has taken place, the rate at which a given protein is
produced may be further affected by the presence or absence of other
proteins in the cell or other factors in the chemical environment.
One way to approach the study of this complicated system of
interacting molecules is to think of each gene, together with its
promoter region, as an agent that interacts with other genes via
protein-mediated interactions.

The logical structure of systems of many agents that exert causal
influences on each other may be represented as a graph in which nodes
represent agents and directed edges indicate the presence of a causal
influence of one agent on another.  In modeling the logic of the cell,
for example, we may think of each node as representing a gene and each
directed link as indicating that the concentration of the protein
produced by one gene has some effect on the production rate of the
other gene's protein.  More precisely, a node represents a gene
together with its promoter region.  An incoming link indicates that
under some circumstances a given protein can affect the transcription
rate of a gene.  The full set of genes (with promoter regions) and
interactions forms a graph that may be called the genetic regulatory
network.  The {\em dynamics} of the network is determined by
parameters (reaction rates) associated with the links and by a
function for each gene that determines the rate of protein production
as a function of the concentrations of its regulators.  

Genetic regulatory networks have, at least along some branches of
evolution, grown over time from a relatively small ancestral genome to
a vastly complex network of over $20,000$ genes.  We would like to
know which features of real genetic regulatory networks are
reflections of simple physical or mathematical laws as opposed to
finely tuned solutions of specific problems faced during the
evolutionary process.  As a first step, we study a class of simple
models of the growth process to see which network features arise
purely from probabilistic effects when selection plays a minimal role.
After discussing the general trends associated with the variation of
certain parameters in the model, we compare the structures generated
by the model to biological data and highlight some problematic issues
in the interpretation of the data.

There is substantial evidence for the hypothesis that network growth
occurs primarily via the duplication of genes and subsequent mutations
of one or both members of the duplicate pair. \cite{genes} Under such
mutations (or imperfect copying) the input and output links to a gene
may not all be preserved as both the promoter region and the protein
produced are altered.  In addition, there is the possibility that a
mutation in a gene will cause its protein to bind to a new promoter
region or protein complex and thereby form a link that could not have
been inherited during duplication alone.  Similarly, a new link could
be formed due to the mutation of a promoter region to a configuration
that now binds a new protein.  We refer to links that are generated by
mutation as ``innovated'' links; those created via duplication are
called ``inherited.''  Studies of yeast and E.~coli genomes suggest
that innovation may account for as much as 50\%\ of the links in the
regulatory network.  \cite{teichmann}

In modeling the growth of a genetic regulatory network one must
incorporate some assumptions about the effect of natural selection.
We make several simplifying assumptions concerning separations of time
scales.  As we are interested here in the evolution of the network
structure that occurs over many generations, we ignore the time scale
corresponding to cellular processes and the lifetime of individual
organisms.  Our evolutionary model assumes three additional scales.
First, there is the typical time required for a duplicated gene to
drift via mutation to a new stable gene.  Call this the ``mutation''
time scale.  Second, we assume a much longer time scale required for
the occurrence of duplication events.  That is, we assume that after a
duplication event whatever mutation is going to occur in the
duplicated gene happens before any other duplication event occurs.
Finally, once a gene has mutated a certain amount and thereby found a
niche for itself in the cell, natural selection is assumed to keep it
stable over time scales long compared to the duplication time scale.
Though this is somewhat of a caricature of evolutionary processes at
the genetic level, it has the virtue of conceptual clarity.  We note
that this model applies only to duplication events that ultimately
lead to an increase in the size of the genome.  It does not attempt to
model duplications that lead to adaptive radiation and eventual
selection of a single duplicated gene as the fittest \cite{francino}.

Consistent with the above assumptions, our networks grow via
duplication/mutation events.  When a gene $G$ (together with its
promoter region) is duplicated to create $G'$, it is assumed that $G$
remains fixed while $G'$ mutates (or that a portion of it is not
copied faithfully) so that some of the duplicated input and output
links at $G'$ cease to function.  The parameters in our growth model
are the probabilities of retaining inherited links after mutation and
the probabilities of innovating links.  Note that the binding of a
protein to a promoter region of DNA breaks the symmetry between input
and output inheritance.  In the output case, the issue is whether a
mutation causes changes in a protein that significantly decrease its
binding affinity to unchanged portions of DNA (or perhaps to other
proteins).  In the input case, the issue is whether the mutation in
the promoter region decreases the binding affinity of a protein that
has not changed.

We consider three classes of models: (1) partial duplication, in which
only a subset of links is inherited during a growth event; (2) partial
duplication with constant probability innovation, in which innovation
probabilities are independent of the characteristics of the candidate
nodes to be linked; and (3) partial duplication with
``rich-gets-richer'' innovation, in which nodes with more inputs have
higher probabilities of forming innovative input links.  The set of
models we study includes as special cases several models studied
previously.
\cite{doro2,doro1,barabasi2,sole,redner,krapivsky,ispolatov1,ispolatov2}
In the present work we emphasize the importance of different
probabilities for input and output inheritance and focus on features
of finite networks rather than scaling laws for arbitrarily large
networks.

For each case we study numerically the statistical features of
networks of up to 2000 nodes grown from seeds with 10 nodes or, for
reasons that will become clear later, 100 nodes.  In several cases we
provide theoretical calculations supporting the numerical results.  We
find that certain choices of parameters result in networks that have
input and output degree distributions similar to those reported for
yeast cells and simultaneously match results on the scaling of the
numbers of transcription factors with genome size.  Our results
suggest that a simple process that totally neglects any specific
information about the biological function of individual genes can
produce a realistic network, but only if the process starts from an
appropriate seed.

Before turning to the details of the growth algorithm, we wish to
emphasize the importance of considering the precise meaning of a link
in the network, since this has direct implications for the comparison
to experimental data.  A link between genes could be taken to mean
that the protein product of a gene is a transcription factor that
binds directly to a gene's promoter region.  This interpretation
allows the network to be determined with relatively straightforward
experiments that test for the binding of a given protein to a given
promoter region.  (See, for example, \cite{lee,alonnets}.)  It is also
possible, however, for proteins that fail to bind directly to DNA to
assert regulatory control through the formation of protein complexes
at a promoter region or even through participation in chemical
processes that occur far from the DNA.  For purposes of simulating
genome-wide transcriptional dynamics, all causal relationships between
the expression levels of two genes should be represented as links in
the network.  The difference between model parameters germane to these
two interpretations is discussed in detail below.

\section{Growth of directed networks through partial duplication}

We make the following definitions:
\begin{description}
\item[a family of nodes] is the set of all nodes arising through a
  chain of duplications of a single ancestor in the seed;
\item[a constitutive node] is a regulator that has no inputs.
\item[a regulator] is any node that has at least one output;
\item[an inert node] is a the result of a duplication event in which all
  inputs and outputs are deleted (a constitutive non-regulator);
\item[a transcription factor] is a regulator that has at least one
  output that has been inherited, perhaps through several generations,
  from a seed regulator that has been designated a transcription
  factor.
\end{description}
In biological terms, a constitutive node represents a gene whose
expression level never changes or else changes only in response to
external environmental variables.  An inert node may represent either
a nonfunctional bit of (junk) DNA or a gene that responds to
environmental factors but remains completely independent of the
activity of any other genes.  A transcription factor represents a gene
whose protein binds to DNA, and it is assumed that mutations never
create new transcription factors from other types of regulators.

The partial duplication model is implemented according to the
following procedure.  We define a time step to be the time between
duplication events.  At each time step a gene $G$ is chosen at random
from the network and duplicated, forming a gene $G'$ that has all the
same input and output links as $G$. One of these identical nodes, say
$G'$, is then assumed to mutate.  Each input link inherited from $G$
is independently tested and kept with probability $c_i$, and each
output link with probability $c_o$.  If $G'$ should lose all of
its inputs and outputs, it is considered to have lost all its
function, and thus is removed from the network entirely.  As mentioned
above, there is no physical symmetry requiring $c_i = c_o$.

We can gain some intuition about the effects of the parameters by
studying some limiting cases that permit analytical solutions. The
simplest of these is $c_i = c_o = 1$, the case in which all links are
kept after every duplication event.  In this case all nodes in the
same family have exactly the same set of inputs and outputs.  The
degree distributions will consist of delta functions whose positions
depend on the seed network chosen and the number of duplication events
that have occurred within each family.

For $c_i$ and $c_o$ less than unity, a master equation describes the
evolution of the degree distributions during growth. \cite{redner} Let
$t$ represent the total number of nodes in the network.  Advancing $t$
by one corresponds to a single duplication event.  Let $N_i(t,k)$ and
$N_o(t,k)$ represent the number of nodes at ``time'' $t$ having $k$
inputs and outputs, respectively.  On average, we have
\begin{widetext}
\begin{eqnarray} \label{eq:ntk}
N_i(t+1,k) & = & N_i(t,k) + \frac{1}{t}\sum_{k'=k}^t \binom{k'}{k} c_i^k (1-c_i)^{k'}N_i(t,k')+  \frac{c_o}{t}[(k-1)N_i(t,k-1) - k N_i(t,k)]. \\
N_0(t+1,k) & = & N_o(t,k) + \frac{1}{t}\sum_{k'=k}^t \binom{k'}{k} c_o^k (1-c_o)^{k'}N_o(t,k') + \frac{c_i}{t}[(k-1) N_o(t,k-1)- k N_o(t,k)]. 
\end{eqnarray}
\end{widetext}
The sum on the right hand side of the first equation represents
probability of the addition of a node with $k$ inputs due to the
duplication and subsequent mutation of a node with $k'$ inputs.  The
$c_o (k-1) N_i(t,k-1) / t$ term is the probability that the duplicated
node is one of the $(k-1)$ inputs to a node $K$ and the inherited
output is kept, so that the number of inputs to $K$ is incremented to
$k$.  The last term comes from the possibility of adding a new input
to a node that already has $k$ inputs.  

In the limit of large $t$, the sums in (\ref{eq:ntk}) can be approximated
by
\begin{equation} \label{eq:mfntk}
\sum_{k'=k}^t \binom{k'}{k} c^k (1-c)^{k'}N(t,k') = \frac{1}{c}N(t,k/c),
\end{equation}
where we have assumed that $N(t,k')$ does not vary significantly over
the range of $k'$ values that have an appreciable probability to yield
$k$ inputs after mutation \cite{doro2}. The factor of $1/c$ comes from
the fact that there are $1/c$ values of $k'$ for which $\lfloor
ck'\rfloor = k$.  This approximation allows for rapid iteration of the
master equation, enabling numerical studies of distributions for large
$t$.  (For analytical results on master equations of this type, see
\cite{redner}.)

\begin{figure}[tbp]
   \centering
   \includegraphics[clip,width=0.72\linewidth]{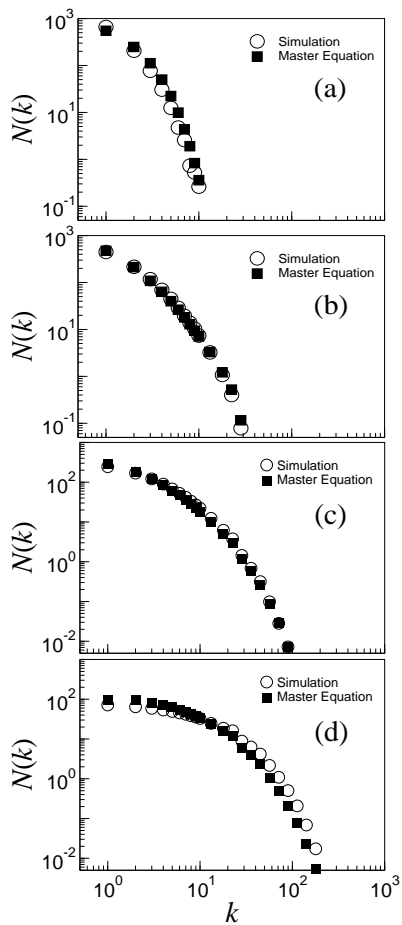}
\caption{Simulated (circles) and predicted (squares) degree
  distribution functions for different values of $c$.  (a) $c =0.1$;
  (b) $c=0.3$; (c) $c = 0.5$; and (d) $c=0.7$.  For models with no
  innovation and $c_i = c_o$ the indegree and outdegree distributions
  are identical.}
\label{fig:agreement}
\end{figure}

For $c_i = c_o \equiv c$ the two master equations become identical and
the asymptotic forms of the input and output distributions are the
same.  For $0.2 \alt c \alt 0.7$, the approximate master equation gives
an accurate estimate of $N(t,k)$ for network sizes $t > 100$, as shown
in Figure~\ref{fig:agreement}.  For $c$ outside this range, however,
the degree distribution predicted by iterating the approximate
equation only matches simulation data for large values of $k$, as
expected.

As $c$ approaches zero, the duplication/mutation process becomes
essentially equivalent to the preferential attachment growth algorithm
\cite{barabasi2} that is known to produce networks with scale free
degree distributions.  The only difference is that the duplication
process produces a large number of inert nodes, whereas preferential
attachment counts only those nodes that do get linked to the existing
network.  For sufficiently small $c$, the probability of producing a
node with more than one input or output via partial duplication is
negligible (on the order of $c^2$), so each new node (that is not
inert) is added to the network with a single input or output.  The
probability of forming a new input to node $G$ is proportional to the
probability of selecting an input node to $G$ for duplication, which
in turn is proportional to the number of inputs that $G$ already has.
For $c\rightarrow 0$, ignoring the production of linkless (inert)nodes
leads to:
\begin{equation} \label{eq:pref}
N(t+1,k) = N(t,k) + \delta(k-1) + \frac{1}{t}[(k-1)N(t,k-1) - k N(t,k)].
\end{equation}
The total number of links in the connected part of the network
approaches $t$ as $c$ gets small, so this master equation is
equivalent to the one given for preferential attachment by Barabasi et
al., which is known to yield a scale-free distribution $N(t,k) =
k^{-\gamma}$ with $\gamma = 3$.  \cite{doro1,barabasi2}

For $c =0.5$, our model is equivalent to one proposed by Dorogovtsev
et al, who predicted scale free behavior in the degree distribution
function, with the frequency of occurrence of indegree (or outdegree)
$k$ decaying like $k^{\sqrt{2}}$. \cite{doro2} This scale free
behavior should occur, however, only over the domain of very large $t$
and $k$.  The systems we have studied are not large enough to show the
predicted scaling, as we can see from the (c) panel in
Figure~\ref{fig:agreement}.  References
\cite{krapivsky,ispolatov1,ispolatov2} also provide analytical results
on the asymptotic scaling properties for small $c$ and failure of
self-averaging for large $c$ in similar models.

\subsection{Effects of different values for $c_i$ and $c_o$}

\begin{figure}[tbp]
\centering
\includegraphics[clip,width=0.6\linewidth]{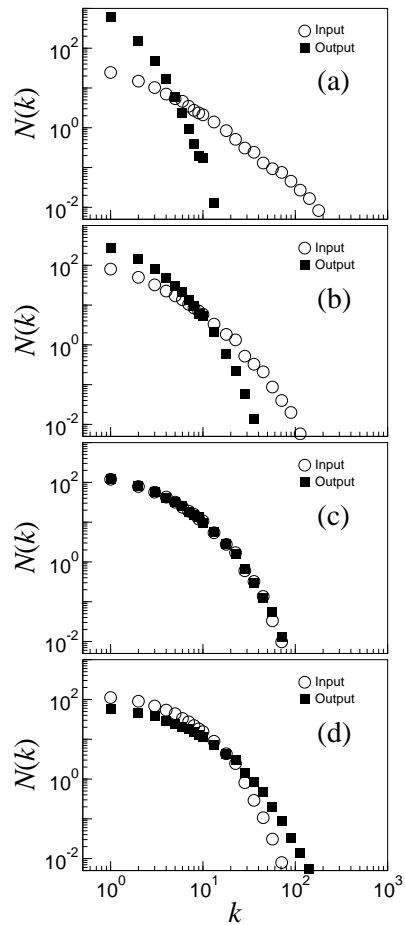}
\caption{Indegree (circles) and outdegree (squares) 
  distribution functions for $c_o = 0.5$ and different values of
  $c_i$: (a) $c_i = 0.1$; (b) $c_i = 0.3$; (c) $c_i = 0.5$; and (d)
  $c_i = 0.7$.  Each data point represents the average number of nodes
  with a given number of inputs in an ensemble of 100 networks.}
\label{fig:cdiff}
\end{figure}

Holding $c_o$ constant at $0.5$, it is interesting to see what happens
as $c_i$ is varied.  To compare the distributions for different
parameter values we choose to keep the total number of (non-inert)
nodes fixed at 1000.  That is, we simulate the network growth (not the
approximate master equations) from a seed of $10$ nodes with $10$
randomly assigned connections, discarding inert nodes and stopping
when the network contains 1000 nodes.  The data shown in
Fig.~\ref{fig:cdiff} are averages taken over 100 networks.

Over the range of sizes probed, the indegree distributions are roughly
scale-free for smaller values of $c_i$ as seen in
Fig.~\ref{fig:cdiff}(a).  We note that the scaling exponent suggested
by these data is approximately $\gamma = 1$, but the systems are not
large enough to be described by the asymptotic scaling regime of the
relevant master equation, so we do not expect this scaling to persist
to substantially larger system sizes.  A surprising result is that
decreasing $c_i$ actually tends to {\em broaden} the tail of the
indegree distribution.  One may have expected a shift of the
distribution favoring smaller indegrees because the probability of
keeping a large number of inputs in a given duplication event
decreases with decreasing $c_i$.  There is a competing effect,
however, associated with the increased production of inert nodes,
which are not counted as part of the 1000 nodes that constitute the
final network.  Because nodes with small indegrees tend to produce
inert nodes, the probability of a node being duplicated is skewed
toward nodes with high indegrees.  In the limit of very small $c_i$,
the situation is similar to the preferential attachment limit
discussed above, where nodes with more inputs are more likely to
receive new inputs when another node is duplicated.  The difference
here is that the newly created node may have many outputs rather than
the fixed number stipulated in standard preferential attachment
models.

Note that Fig.~\ref{fig:cdiff}(c) is not identical to
Fig.~\ref{fig:agreement}(c), even though they correspond to the same
parameter values.  This is because of the difference in the criteria
used for determining the sizes of the networks in the ensemble.  For
purposes of comparing distributions to those predicted by the master
equation, we must include inert nodes in the simulation.  That is,
inert nodes are counted as contributing to the system size.  For the
purpose of examining an ensemble of networks with a given number of
genes, however, we do not count the inert nodes.  Interpreted in this
context, the data shown in Fig.~\ref{fig:agreement}(c) correspond to
an ensemble with a distribution of networks sizes, all smaller than
1000.

\section{Effects of Innovation}

The pure partial duplication model does not contain a mechanism for
innovation of new links.  We now consider two models for representing
the general effects of innovation.  The first model is similar to that
one introduced by Sol\'{e} \cite{sole}: each time a new node is
created, every possible input and output link it might form is tested
and kept with independent probability $p$.  In the second model, the
probability of keeping a tested link is assumed to depend on the
indegree of the node at the input side of the link; nodes with more
inputs are assumed to have a higher probability of accepting new
inputs.  The motivation for the latter comes from the ability of
proteins to form complexes or interact in ways that affect
transcription, which suggests that if more proteins participate in the
regulation of a given gene, there are more opportunities for a new
protein to exert a regulatory influence.  Note that this is not true
on the output side; the probability that a given protein will
participate in a particular regulatory link should not change just
because that protein begins to participate in additional links.

\subsection{Constant probability innovation}

Innovation is added to the partial duplication model as follows.  At
each time step a gene $G'$ is produced by partial duplication.  $G'$
is then given a chance to develop an innovative input from all of the
transcription factors in the network.  Note that in this model all
regulators are transcription factors.  It is impossible for a node
with no outputs to become a regulator.  If gene $G'$ has inherited
outputs (and thus is a transcription factor), then it is given a
chance $p$ to bind to every node in the network.  If $G'$ has no
inputs or outputs after all inherited and innovated links are tested,
it becomes inert and is assumed to remain inert forever; it can no
longer receive innovated inputs.

\begin{figure}[tbp]
\centering
\includegraphics[clip,width=0.6\linewidth]{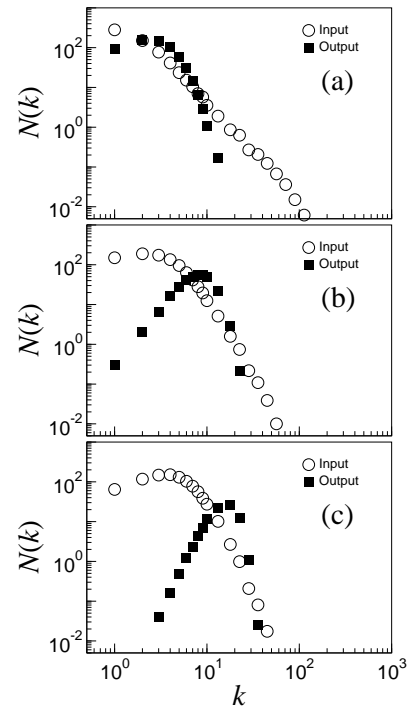}
\caption{Indegree (circles) and outdegree (squares)
  distribution functions for $c_o = 0.5$ and $c_i = 0.1$, with various
  values of $p$: (a) $p = .001$; (b) $p = .005 $; (c) $p = .01 $.
  Each different value of $p$ results in a different percentage of
  innovated links: (a)$39.1$\% (b) $59.8$\% (c)$64.6$\%.}
\label{fig:constp1}
\end{figure}

\begin{figure}[tbp]
  \includegraphics[clip,width=0.6\linewidth]{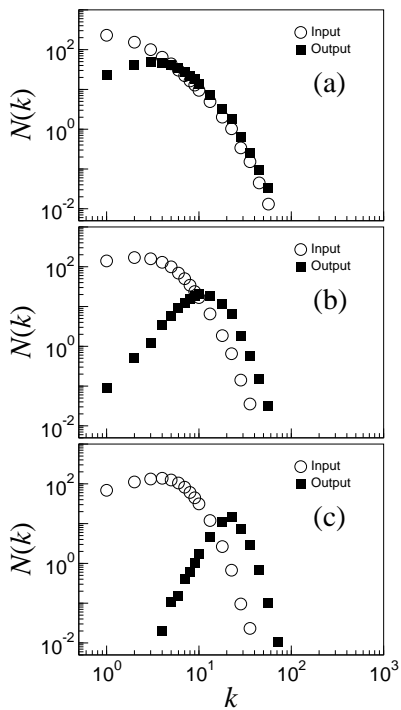}
\caption{Indegree (circles) and outdegree (squares) 
  distribution functions for $c_o = 0.5$ and $c_i = 0.5$, with various
  values of $p$: (a) $p = .001$; (b) $p = .005 $; (c) $p = .01 $.
  Each different value of $p$ results in a different percentage of
  innovated links: (a)$16.2$\% (b) $36.5$\% (c)$43.2$\%.}
\label{fig:constp3}
\end{figure}

For the present study, networks were grown until the number of
non-inert nodes reached $N = 1000$.  Figs.~\ref{fig:constp1} and
\ref{fig:constp3} show the indegree and outdegree distributions for
$c_o = 0.5$ and $c_i = 0.1$ and $0.5$, respectively, for several
values of $p$.

%\begin{figure}[tbp]
%\includegraphics[clip,width=0.6\linewidth]{ProjectNov350011000.eps}
%\includegraphics[clip,width=0.6\linewidth]{ProjectNov350051000.eps}
%\includegraphics[clip,width=0.6\linewidth]{ProjectNov35011000.eps}
%\caption{These plots show the indegree (circles) and outdegree (squares) distribution functions for $c_o = 0.5$ , $c_i = 0.3$, and various values of $p$ (a) $p = .001$; (b) $p = .005 $; (c) $p = .01 $; The x-axis shows values of $k$ and the y-axis the average number of nodes with that $k$. Each different value of $p$ results in a different percent of links innovated: (a)$25.6$\% (b) $47.7$\% (c)$53.5$\%.}
%
%\label{fig:constp2}
%\end{figure}
%

Comparing Figs.~\ref{fig:constp1} and \ref{fig:constp3} to
Fig.~\ref{fig:cdiff}, we see that innovation serves to shift the
weight in the outdegree distribution to larger values of $k$ and
produces a peak at nonzero $k$.  This qualitatively matches the yeast
data of Lee et~al., who find that the output distribution has a peak
at a value near $30$.\cite{lee} However, innovation also {\em reduces}
the weight in the tail of the indegree distribution.  This somewhat
surprising result is due to the fact that innovation decreases the
number of nodes with only a single input, thereby decreasing the rate
at which duplications yield inert nodes, which in turn decreases the
number of duplications of high indegree nodes.

\subsection{Rich-gets-richer innovation}

As mentioned above, links corresponding to the direct bindings of
transcription factors to DNA do not exhaust the possible sources of
regulatory control.  To model the networks observed in experiments
that detect the influence of protein interactions and other regulatory
effects, one must consider models in which non-regulatory nodes can
develop outputs via innovation.  If protein interactions are an
important source of regulatory control, we might expect that any gene
could possibly innovate an output to any other gene, and that genes
with more inputs would be more likely to gain further inputs.  This
``rich-gets-richer'' effect may produce scale free indegree
distributions, as in simpler preferential attachment models.

One might imagine many different ways in which the probability of
receiving an innovated input could depend on the indegree of a node.
The situation of interest, though, is one in which the interactions
between nodes are determined by physical properties of molecules and
therefore should not depend on the order in which nodes were added to
the system.  That is, the probability that a link from $G$ to $H$ is
innovated may depend on the number of inputs to $H$, but should not
depend on whether those inputs were generated before or after $G$ was
created.  This constraint restricts the class of models considerably,
as described below.

The model is implemented according to the following procedure.  At
each time step, first a gene $G$ is chosen at random from the network
and partial duplication occurs with probabilities $c_i$ and $c_o$ as
above.  Each gene in the network is then tested to see if it innovates
an input to $G'$ and/or receives an innovative input from $G'$.
Whenever a new gene $G'$ is created, it is given a probability $p(n)$
of having an output to each gene $H$, where $n$ is the number of
inputs $H$ already has.  (Note that $p(0)$ does not vanish.  There is
some probability that the new gene will regulate a gene that
previously had no inputs.)  In addition, any time that any gene $J$
acquires one or more new inputs, whether via duplication or
innovation, all other genes $K$ in the network are given probability
$q(\Delta)$ of forming a new input link to $J$, where $\Delta$ is the
number of inputs that $J$ has gained since the last time the formation
of an innovative link from $K$ to $J$ was tested.  This latter process
is iterated until a complete pass through the network generates no new
links.  Multiple time steps are performed until the network contains
$N$ genes with at least one input or output link.

To ensure that the probabilities of links being present are
independent of the order in which new innovations are attempted, we
must choose
\begin{eqnarray}\label{eq:pn}
p(n) & = & 1 - e^{-(1+n)/n_0}; \\
q(\Delta) & = & 1 - e^{-\Delta/n_0},
\end{eqnarray}
where $1/n_0$ gives the probability that a gene with no inputs will
obtain an input from any particular gene.  The exponential form of
$q(\Delta)$, means that $[1-q(\Delta_1)][1-q(\Delta_2)] =
[1-q(\Delta_1+\Delta_2)]$ for any $\Delta_1$ and $\Delta_2$.  This
ensures that the probability of there being no link between two genes
is independent of the number of times the potential link was tested
during the growth process.  Thus our model is a self-consistent growth
algorithm in which the probability of adding a new link can always be
determined without worrying about the order in which links are tested
and added to the system.

\begin{figure}[tbp]
\centering
\includegraphics[clip,width=0.6\linewidth]{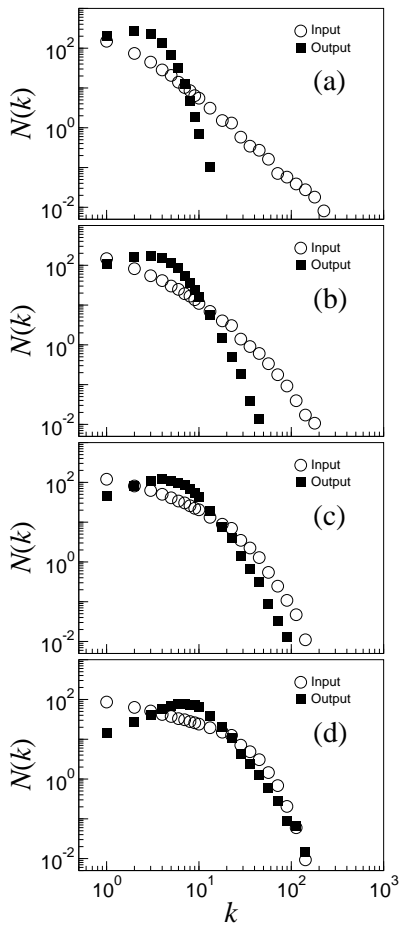}
\caption{Indegree (circles) and outdegree (squares)
  distribution functions for $c_o = 0.5$ and different values of
  $c_i$: (a) $c_i = 0.1$; (b) $c_i = 0.3$; (c) $c_i = 0.5$; and (d)
  $c_i = 0.7$, with $n_0 = 3000$ in the rich-gets-richer innovation
  model. The percentage of innovated links is (a) $48.3$\%; (b)
  $41.4$\%; (c) $37.5$\%; (d) $35.3$\%.}
\label{fig:richer}
\end{figure}

The amount of innovation and the values of $c_i$ and $c_o$ are roughly
similar between Fig.~\ref{fig:constp1} (a) and Fig.~\ref{fig:richer}
(a), Fig.~\ref{fig:constp3} (b) and Fig.~\ref{fig:richer} (c).  These
pairs of figures show that rich-gets-richer innovation broadens the
indegree distribution compared to constant probability innovation, as
expected when high indegree nodes are favored for innovated inputs.
We also note the general trend that rich-gets-richer innovation gives
rise to a less pronounced peak in the outdegree distribution and to
clearer power law tails.

The differences between constant probability innovation and
rich-gets-richer innovation may be relevant for the modeling of
biological data.  At the very least, these differences highlight the
importance of obtaining a clear understanding of the types of
regulatory interactions that are included in experimental reports on
the structure of genetic regulatory networks.

\subsection{Selecting a starting seed}

The discussion and results above focused on generic behaviors expected
when a network has grown to many times the size of the initial seed.
Unfortunately, it appears that no choices of the parameters $c_i$,
$c_o$, and $p$, can reproduce degree distributions qualitatively
similar to those reported for yeast in experiments measuring
transcription factor binding.  In those experiments, it is found that
the output distribution is extremely broad, including an appreciable
number of nodes with more than 130 outputs, while the input
distribution remains quite narrow, containing very few nodes with more
than about 10 inputs. \cite{lee,harbison}

There is, however, another piece of biological evidence that suggests
that real genetic regulatory networks cannot be in the asymptotic
large $N$ regime.  Nimwegen has observed that the number of
transcription factors in an organism scales like $N^\alpha$, with
$alpha\approx 1.26$ for eukaryotes and $\alpha\approx 1.87$ for
bacteria, where $N$ is the size of the genome. \cite{nimwegen} A
scaling law of this type with $\alpha$ greater than unity is
impossible for arbitrarily large genome sizes since the number of
transcription factors cannot exceed the total number of genes.  We are
thus led to consider models in which the number of transcription
factors is initially quite small compared to the total number of nodes
in the network.

We study networks of 2000 nodes grown from a seed consisting of a
single node that regulates itself and 99 other non-regulating nodes.
This is roughly consistent with an extrapolation from Nimwegen's
observations to genomes with only 100 genes.

\begin{figure}[tbp]
  \includegraphics[clip,width=0.6\linewidth]{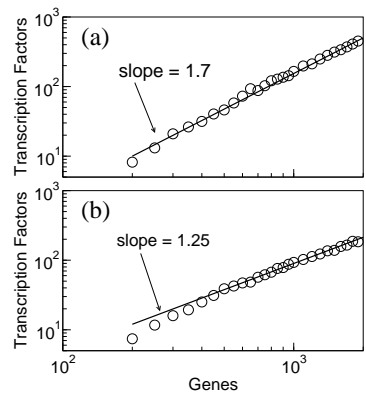}
\caption{Growth rate of the number of transcription factors for the 
  parameters $c_i = 0.2$ , $c_o = 0.5$, with (a) $p=0$ and (b)
  $p=0.005$.  The networks are grown from a seed of one regulator
  linked to 99 other nodes.  Solid lines show the scaling laws
  consistent with Nimwegen's analysis of (a) bacteria and (b)
  eukaryotes.}
\label{fig:tfscaling}
\end{figure}

Since we are now considering data that identify transcription factors
rather than all regulatory interactions, we work with the constant
probability innovation model.  As seen in Fig.~\ref{fig:tfscaling},
the number of transcription factors does indeed grow roughly as a
power law, with an exponent that depends on the value of $c_o$ and
$p$.  The straight lines shown on the plots indicate scaling exponents
roughly consistent with Nimwegen's reported values.  (No attempt was
made to search parameter space for optimal fits to Nimwegen's exponent
for bacteria.)

In Fig.~\ref{fig:tfscaling}, one can see evidence for a region of
quadratic scaling for very small system sizes, implying that when a
gene is selected for duplication, the result is twice as likely to be
kept in the genome if it is a transcription factor than if it is not.
\cite{nimwegen} We can understand this effect in the context of our
model as follows.  If $c_o$ is sufficiently large and transcription
factors have many outputs, the chance that partial duplication of a
transcription factor will result in an inert node is negligible.  On
the other hand, most genes have rather few inputs, so the probability
of duplication of a non-transcription factor leading to an inert node
is appreciable and is also sensitive to $c_i$.  Since duplications of
transcription factors increase the number of inputs to many of the
nodes, exact calculation of the rate at which inert nodes are
generated is difficult.  We find numerically that $c_i \approx 0.2$
leads to an initial growth phase with roughly quadratic scaling.

\begin{figure}[tbp]
\includegraphics[clip,width=0.6\linewidth]{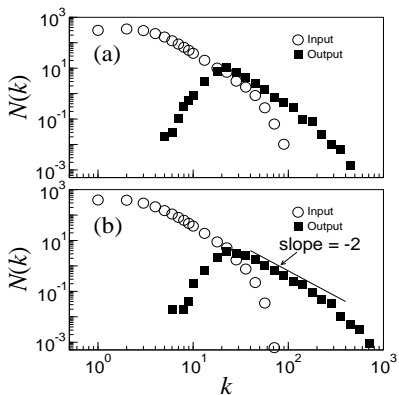}
\caption{(a) Indegree (circles) and outdegree (squares) 
  distribution functions for the model parameters equal to those of
  Fig.~\ref{fig:tfscaling}(b): $c_o = 0.5$, $c_i = 0.2$ and $p=0.005$.
  (b) Same as (a) except $c_i = 0.3$.  The networks are grown from a
  seed of one regulator linked to itself and 99 other nodes.}
\label{fig:yeast1}
\end{figure}

It is interesting to compare the degree distributions obtained from
the model with parameters that yield a scaling exponent consistent
with Nimwegen's analysis of eukaryotes to the distributions reported
for yeast. \cite{lee} Fig.~\ref{fig:yeast1} shows that we obtain an
indegree distribution with no extended power-law tail together with an
outdegree distribution with a broad power-law tail.  The widths and
shapes of the distributions are roughly consistent with those observed
for yeast, including such features as the power-law outdegree tail
with exponent near $2$ and a slower-than-exponential decay at small
indegree that is rapidly cut off above a maximum near 30 in the
outdegree distribution.  For the model parameters used here,
approximately 28\%\ of the links in the system are formed via
innovation.

We note that Harbison et al. \cite{harbison}, using a different
approach from Lee et al. \cite{lee}, report data on outdegrees for 203
transcription factors that show a distribution extending to
approximately 300, but appearing to have an exponential form and no
peak.  In the Harbison data, however, outdegrees are determined as the
union of results of different preparations that show wide variations
in the outdegrees of individual genes.  A third source of data on the
yeast network is available from Milo et al. \cite{alonnets}
Careful comparative analysis of all of the available data is beyond
the scope of the present work.

\section{Conclusions}

We have introduced and studied two classes of growth algorithms for
networks with directed links.  Both involve partial duplication of
nodes as the sole growth mechanism and separate parameters determining
the probabilities of keeping inputs and outputs of the duplicated
node.  In one case, the opportunities for the innovation of new links
are given a constant probability and every possible output link is
tested exactly once during the growth.  In the other, nodes with more
input links are given a higher probability of receiving new inputs,
and potential input links are re-tested every time a node receives a
new input.  In the context of our simplified description of genome
growth, the first model is appropriate for studying transcription
factor binding networks, where output links correspond only to direct
binding of a protein to DNA.  The second model corresponds to the full
network of regulatory interactions, in which some proteins that do not
bind to DNA can still exert regulatory control.

We have studied indegree and outdegree distributions and the relation
between the number of transcription factors and system size in
networks of up to 2000 nodes grown from random seeds with 10 nodes or
special seeds with 100 nodes.  Some counterintuitive results
concerning the effects of various parameters on the degree
distributions were observed and explained.  Parameters were found that
produce networks with degree distributions similar to those reported
for yeast cells and plausibly realistic scaling laws for the number of
transcription factors as a function of genome size in eukaryotes.

For our growth model, the production of realistic networks requires a
seed in which many nodes are regulated by a single transcription
factor with a self-input.  While we do not know the detailed
regulatory network structure of any ancestral organism with only 100
genes, it is plausible to suggest that early simple organisms required
relatively few regulatory genes that controlled a large number of
structural genes.

Several features of transcriptional network architecture would be
natural candidates for further study.  We have not yet analyzed the
frequency of occurrence of small network motifs or clustering
statistics.  It is known that partial duplication induces strong local
correlations in a model with identical probabilities for retaining
input and output links. \cite{sole} We conjecture that minor
modifications to our model favoring innovation between genes that
share a neighbor could account for higher clustering coefficients and
favor local motifs without altering the overall degree distributions
or transcription factor scaling laws.

We have explored the behavior of a class of network growth models that
may be relevant for understanding the structure of genetic regulatory
networks.  Our results indicate that several nontrivial features of
presently available biological data can arise simply from
probabilistic growth rules with no notion of optimization due to
selection coming into play.  This suggests that statistical features
of real genetic regulatory networks may be determined by physical or
biochemical parameters rather than careful tuning through natural
selection.

This is not to say that natural selection plays no role in determining
which networks survive and prosper.  Among the ensemble of networks
generated by our growth models, there are many possibilities for
variation.  For example, the causal influences indicated by links in
our networks, and hence the dynamical properties of the network, may
evolve under selection pressures via small mutations that do not
affect the network architecture.  Nevertheless, it is interesting to
see that simple probabilistic growth rules can account for an ensemble
of possibilities available to the fitness selection process.
\cite{kauffmanensemble}

\acknowledgments{S.~K.~thanks the Department of Cell Biology and
  Physiology at the University of New Mexico, Albuquerque, for support
  during the early stages of this work.  This material is based upon
  work supported by the National Science Foundation under Grant No.
  PHY-0417372.}

\bibliographystyle{apsrev}
\bibliography{rbn}

\end{document}